\documentclass{article}
\usepackage{graphicx} % Required for inserting images
\usepackage{multirow}
\usepackage[table]{xcolor}
\usepackage{amsmath}
\usepackage{amsfonts}
\usepackage{enumitem}
\usepackage{float}
\usepackage{optidef}
\usepackage{subfig}
\usepackage{caption}
\usepackage[mathlines]{lineno}
\usepackage{authblk}
\usepackage[
doi=false,isbn=false,url=false,eprint=false,
backend=biber,
style=numeric,
sorting=none,
giveninits=true,
maxbibnames=3
]{biblatex}
\renewbibmacro{in:}{}
\AtEveryBibitem{%
  \clearlist{language}%
  \clearfield{pages}
  \clearfield{note}
  \clearfield{month}
}
\begin{document}

\title{Interplay-robust optimization for treating irregularly breathing lung patients with pencil beam scanning}

\author[1,2]{Ivar Bengtsson\thanks{Contact: ivarben@kth.se}}
\author[1]{Anders Forsgren}
\author[2]{Albin Fredriksson}
\author[3]{Ye Zhang}

\affil[1]{Department of Mathematics, KTH Royal Institute of Technology, Stockholm, Sweden}
\affil[2]{RaySearch Laboratories AB, Stockholm, Sweden}
\affil[3]{ Center for Proton Therapy, Paul Scherrer Institut, Villigen, Switzerland}

\date{\today}

\maketitle
%\linenumbers

\begin{abstract}

    \textbf{Background}: The steep dose gradients obtained with pencil beam scanning (PBS) allow for precise targeting of the tumor but come at the cost of high sensitivity to uncertainties. Robust optimization is commonly applied to mitigate uncertainties in density and patient setup, while its application to motion management, called 4D-robust optimization (4DRO), is typically accompanied by other techniques, including gating, breath-hold, and re-scanning. In particular, current commercial implementations of 4DRO do not model the interplay effect between the delivery time structure and the patient's motion.
    
    \textbf{Purpose}: Interplay-robust optimization (IPRO) has previously been proposed to explicitly model the interplay-affected dose during treatment planning. It has been demonstrated that IPRO can mitigate the interplay effect given the uncertainty in the patient's breathing frequency. In this study, we investigate and evaluate IPRO in the context where the motion uncertainty is extended to also include variations in breathing amplitude.
    
    \textbf{Methods}: The compared optimization methods are applied and evaluated on a set of lung patients. We model the patients' motion using synthetic 4DCTs, each created by deforming a reference CT based on a motion pattern obtained with 4DMRI. Each synthetic 4DCT contains multiple breathing cycles, partitioned into two sets for scenario generation: one for optimization and one for evaluation. Distinct patient motion scenarios are then created by randomly concatenating breathing cycles varying in period and amplitude. In addition, a method considering a single breathing cycle for generating optimization scenarios (IPRO-1C) is developed to investigate to which extent robustness can be achieved with limited information. Both IPRO and IPRO-1C were investigated with 9, 25, and 49 scenarios.
    
    \textbf{Results}: For all patient cases, IPRO and IPRO-1C increased the target coverage in terms of the near-worst-case (5th percentile) CTV D98, compared to 4DRO. After normalization of plan doses to equal target coverage, IPRO with 49 scenarios resulted in the greatest decreases to OAR dose, with near-worst-case (95th percentile) improvements averaging 4.2 \%. IPRO-1C with 9 scenarios, with comparable computational demands as 4DRO, decreased OAR dose by 1.7 \%.
    
    \textbf{Conclusions}: The use of IPRO could lead to more efficient mitigation of the interplay effect, even when based on the information from a single breathing cycle. This can potentially decrease the need for real-time motion management techniques that prolong treatment times and decrease patient comfort.
    
\end{abstract}

\section{Introduction}

The ability of \textit{pencil beam scanning} (PBS) to accurately shape dose distributions with the use of intensity modulation makes it the preferred delivery technique for proton radiation therapy treatments \cite{lomax_treatment_2004}. However, this accuracy comes at the cost of high sensitivity to uncertainty in parameters affecting the dose delivery, such as tissue density, patient positioning, and patient motion. Robust optimization, which works by simultaneously considering and optimizing with respect to multiple error scenarios, has been clinically implemented to mitigate density- and positioning uncertainties \cite{fredriksson_minimax_2011, janson_treatment_2024, mohan_review_2022}. However, its application to motion management is typically complemented with active and/or passive motion management techniques during the treatment delivery \cite{zhang_survey_2023, knausl_surveying_2023, knausl_review_2024}. Active techniques, such as beam gating (with or without breath-hold), are based on observing and then mitigating (and possibly controlling) the patient's motion. On the other hand, passive re-scanning techniques mitigate motion by dividing the delivery of the plan into multiple sub-deliveries, averaging out the dose degradation effects. All of the aforementioned motion management techniques suffer from drawbacks, either through prolonged treatment times or decreased patient comfort.

Current clinical implementations of robust optimization for motion mitigation, often called \textit{4D-robust optimization} (4DRO), extend the scenario definition from including a density scaling and a positioning shift to also including an anatomical state, typically represented by a 4DCT motion phase. The dose in a scenario is then computed under the assumption that all beam spots are delivered to a single phase. Consequently, a limitation of 4DRO is that it does not consider the interplay effect, which is the interference between the patient's motion and the time structure of the delivery \cite{lambert_intrafractional_2005, seco_breathing_2009, bert_quantification_2008, bert_motion_2011}. Therefore, optimization models that consider the interplay effect are of research interest.

Previously, Bernatowicz et al.\ have proposed an optimization method that used \textit{4D dose computations} (4DDCs) in the optimization, explicitly accounting for the interplay effect \cite{bernatowicz_advanced_2017}. Hereafter, we refer to this use of 4DDC in optimizing PBS plans as \textit{interplay-driven optimization} (IPO). Similarly, Graeff et al.\ have proposed a system that performs IPO and then uses the treatment control system to ensure the precise delivery of each pencil beam spot to the appropriate motion phase \cite{graeff_multigating_2014}. The main challenge of IPO is analogous to the general uncertainty problem in PBS: failure to accurately represent the actual patient motion on the day of treatment leads to discrepancies between the intended and the realized treatment outcomes.

As a remedy to the problem of patient motion uncertainty, Engwall et al.\ have proposed \textit{interplay-robust optimization} (IPRO) \cite{engwall_4d_2018}. This approach includes multiple patient motion scenarios in the optimization to represent the motion variability during the delivery. The study showed that IPRO can maintain robust target coverage to a greater extent than 4DRO. A limitation of their study was the reliance on a single pre-treatment 4DCT used to model breathing motion in the numerical experiments. Consequently, the generated evaluation motion scenarios were based on only ten motion states per patient, varying in the period of each breathing cycle. Therefore, although IPRO was shown to mitigate uncertainty in patient breathing frequency, its ability to mitigate uncertainty related to variation in breathing amplitude remains to be demonstrated.

In this work, we extend the investigation of IPRO to the case of irregular breathing, in which the patient sequentially passes through distinct breathing cycles, varying in frequency and peak amplitude. Our breathing model is based on motion states that were previously synthesized in work by Duetschler et al.\ \cite{duetschler_synthetic_2022}, which used a reference CT and 4DMRI to generate synthetic 4DCTs (s4DCT), spanning multiple breathing cycles varying in period and amplitude, with deformable image registration. The optimization method used considers doses computed by 4DDC using phase sorting \cite{bert_quantification_2008, grassberger_motion_2013, engwall_effectiveness_2018, engwall_4d_2018} extended to irregular motion. For evaluation, the delivered doses are computed analogously on breathing data unseen at the optimization stage. This way, we aim to answer if IPRO can provide dosimetric advantages over 4DRO in the case of irregular breathing, where the patient's motion during delivery is only partially known during the treatment plan optimization.

\section{Method}

This work uses a time-varying volumetric image to represent the patient's motion. In particular, this image is represented by a s4DCT. A snapshot of the s4DCT at a particular time is called a \textit{motion state}. Our 4DDCs follow a phase-sorting-based approach, in which partial doses are computed on each motion state depending on the exact delivery times of each pencil beam spot and accumulated on a reference state. We then perform interplay-driven optimizations considering doses computed with 4DDC.

\subsection{4D dose computation} \label{4DDC}

Consider an index set $\mathcal{P}$ that enumerates all plausible motion states of the patient. A plausible motion pattern of the patient during the delivery of a treatment fraction is called a \textit{motion scenario} and is described by a function $s(t)$ that maps time points to motion states, $s: \mathbb{R}_+ \to \mathcal{P}$.

Then consider the set $\mathcal{I}^p(x; s)$ as the indices of the spots delivered on state $p$ given for the spot weights $x$
in scenario $s$. Given the dose influence matrix $D^p$ that encodes in its columns the dose of each pencil beam as if delivered during state $p$, the partial dose $d^p(x; s)$ can be computed considering only the relevant columns of $D^p$:

\begin{equation}
    d^p(x; s) = D^p \Tilde{x}^p(x; s),
\end{equation}

\noindent where the auxiliary variable $\Tilde{x}^p(x; s)$ is defined from

\begin{equation}
    \Tilde{x}_i^p(x; s) = \begin{cases}
        x_i \quad \text{if } i \in \mathcal{I}^p(x; s), \\
        0 \quad \text{otherwise.}
    \end{cases}
\end{equation}

\noindent The fraction dose is then the accumulation of the deformed partial doses:

\begin{equation}
    d(x; s) = \sum_{p \in \mathcal{P}} R^p d^p(x; s),
\end{equation}

\noindent where each $R^p$ is a matrix---resulting from deformable registration---that encodes the dose deformation from a state $p$ to the reference state.

\subsection{Interplay-robust optimization}

The main principle of IPRO is to minimize an objective function that takes as input the doses computed with 4DDC considering scenarios in a set $\mathcal{S}$:

\begin{mini}|l|
{x \in \mathcal{X}}{\underset{s \in \mathcal{S}}{\text{max} \quad} f(d(x; s)),}{}{}
\label{IPRO}
\end{mini}

\noindent where $\mathcal{X}$ is the set of feasible spot weights. Here, we recognize that in the typical case of the delivery time structure for PBS, the 4D-computed dose $d(x; s)$ is a non-linear and non-convex function in $x$, stemming from the fact that weights of earlier spots will influence the start time of later spots. However, as previous studies have successfully mitigated this non-linearity heuristically by keeping the set $\mathcal{I}^p(x; s)$ fixed in $x$ and updating it regularly during the numerical optimization process, we will employ the same strategy in this work \cite{bernatowicz_advanced_2017, engwall_4d_2018}. Importantly, IPO can then be seen as a special case of Problem \eqref{IPRO} with $\mathcal{S}$ as a set containing a single scenario.

\subsection{Data}

Our 4DDCs depend on the time structure of the PBS delivery, the considered motion states, and the order in which they appear in each motion scenario.

\subsubsection{Delivery time structure} \label{time_structure}

The delivery time structure was given by a model that describes the dose rate, the lateral scanning time, and the energy switching time -- all based on fits of the experimental data from an IBA ProteusPlus system (IBA, Louvain-La-Neuve, Belgium) in Pfeiler et al.\ \cite{pfeiler_experimental_2018}. The dose rate, measured in milliseconds per monitor unit, was replicated exactly as the model fit in their Figure 6. The lateral scanning speeds were approximated by a linear fit of the data points in their Figure 7. Finally, the energy switching time was a constant 1230 ms.

\subsubsection{Synthetic 4DCT generation}

The s4DCTs employed in this study were previously generated in the work by Duetschler et al.\ \cite{duetschler_synthetic_2022}. The method works by deforming a reference CT with the deformable vector fields produced by the deformable registration of motion acquired with 4DMRI at 2.25 Hz. With a 4DMRI of sufficient duration, one can generate enough motion states to cover the duration of a PBS treatment. For a more detailed description of the s4DCT used, the reader is referred to the original paper \cite{duetschler_synthetic_2022}.

\subsubsection{Patient cases and motion patterns}

\begin{figure}
    \centering
    \makebox[\textwidth][c]{
    \includegraphics[width=1.4\textwidth]{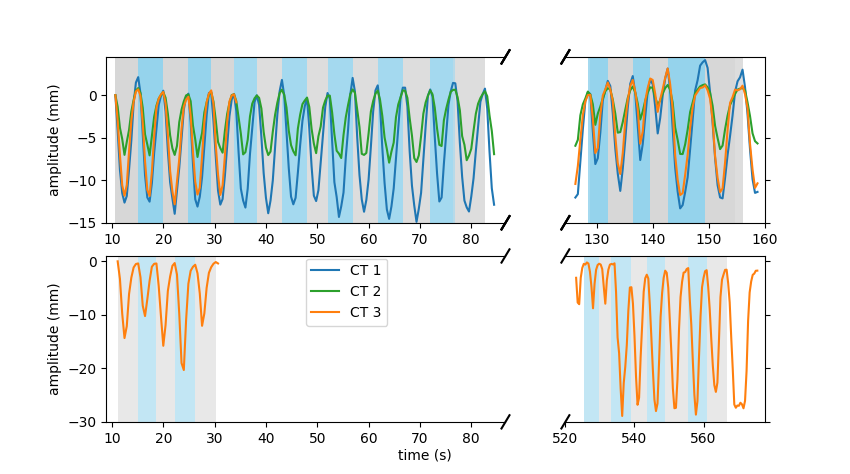}}
    \caption{The motion patterns (amplitude (mm) over time (s)) in the four cases. The motion pattern from 4DMRI 1 (top) is considered for all three reference CTs, while that from 4DMRI 2 (bottom) is considered only for CT 3. The background is shaded to distinguish individual breathing cycles and to indicate the partitioning of the set of breathing cycles into one subset for optimization (odd indices/ gray) and one for evaluation (even indices/ blue).}
    \label{motion}
\end{figure}

The investigated patient cases involved three unique reference CTs (CT 1, 2, and 3) of non-small cell lung cancer patients and two different breathing patterns collected with 4DMRI from healthy volunteers (4DMRI 1 and 2). We considered four distinct cases for use in the numerical experiments; the first three (case 1, case 2, and case 3a) used each CT combined with 4DMRI 1, and the last one (case 3b) used CT 3 with 4DMRI 2. Although the same 4DMRI was used for cases 1, 2, and 3a, the motion magnitudes vary. The four motion patterns are shown in Figure \ref{motion}. For all cases, the motion was predominantly in the inf-sup dimension. Therefore, we estimate the motion magnitude as the inf-sup-coordinate of the deformation vector in the center of the CTV.

\subsection{Scenario modeling} \label{scenario_modeling}

After identifying the peaks of the motion signal, we defined distinct breathing cycles by grouping the motion states between neighboring peaks (corresponding to the end-exhale phase). This process produced a set of breathing cycles per patient, each comprising 7--15 motion states. The purpose of this grouping was to allow for the generation of a large and varied set of plausible breathing scenarios per patient. Since the motion patterns were acquired at a constant rate of 2.25 Hz, the scenario function $s$ was implicitly defined by an ordered sequence of motion states of sufficient duration to cover the treatment delivery.

We then partitioned the set of breathing cycles into two subsets: one for plan optimization and the other for evaluation of plan delivery. Both motion patterns consisted of sub-patterns acquired during distinct time intervals. Therefore, the partition was made to avoid bias toward a specific breathing trend by selecting every other breathing cycle for the optimization set (odd indices) and the rest for the evaluation set (even indices). This selection is visualized by the varying background shades in Figure \ref{motion}, and two resulting sets of breathing cycles are further described in Table \ref{breathing_statistics}.

\begin{table}[h!]
\centering
\makebox[\textwidth]{%
\begin{tabular}{|p{1cm}|p{1cm}|p{1.2cm}|p{1cm}|p{1cm}|p{1cm}|p{1.2cm}|p{1cm}|p{1cm}|}
    \hline
    \multirow{2}{*}{\textbf{Case}} & \multicolumn{4}{c|}{\textbf{Optimization set}} & \multicolumn{4}{c|}{\textbf{Evaluation set}} \\ \cline{2-9} 
    & \textbf{Count} & \textbf{Period (s)} & \textbf{Max.} \newline \textbf{disp.} \newline \textbf{(cm)} & \textbf{Min.} \newline \textbf{disp.} \newline \textbf{(cm)} & \textbf{Count} & \textbf{Period (s)} & \textbf{Max.} \newline \textbf{disp.} \newline \textbf{(cm)} & \textbf{Min.} \newline \textbf{disp.} \newline \textbf{(cm)} \\ \hline
    1 & 11 & $4.85 \pm 0.94$ & $0.13 \pm 0.11$ & $-1.24 \pm 0.27$ & 10 & $4.62 \pm 0.89$ & $0.13 \pm 0.11$ & $-1.22 \pm 0.23$ \\ \hline
    2 & 11 &$4.81 \pm 0.89$ & $0.06 \pm 0.04$ & $-0.63 \pm 0.18$ & 10 & $4.67 \pm 0.89$ & $0.06 \pm 0.04$ & $-0.63 \pm 0.16$ \\ \hline
    3a & 6 & $4.44 \pm 0.68$ & $0.10 \pm 0.06$ & $-0.98 \pm 0.37$ & 5 & $4.44 \pm 1.32$ & $0.13 \pm 0.11$ & $-0.96 \pm 0.27$ \\ \hline
    3b & 7 & $4.57 \pm 1.11$ & $-0.11 \pm 0.10$ & $-1.84 \pm 0.72$ & 6 & $4.74 \pm 0.80$ & $-0.08 \pm 0.08$ & $-2.08 \pm 0.85$ \\ \hline
\end{tabular}
}
\caption{Statistics for each case of the breathing cycles selected for optimization (odd indices) and evaluation (even indices), respectively. Each quantity is expressed using the mean and standard deviation ($\mu \pm \sigma$) over the selected set of breathing cycles.}
\label{breathing_statistics}
\end{table}

\begin{figure}[h!]
\centering
\subfloat[Scenarios sampled with the method used for IPRO and evaluation.]{
\includegraphics[width=\textwidth]{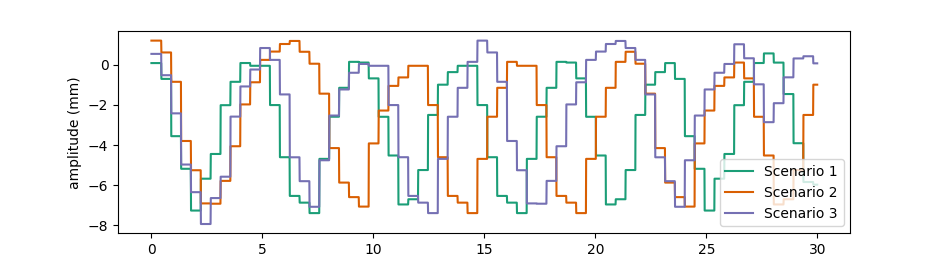}
\label{shuffle_scenarios}
}

\subfloat[Scenarios generated using a single breathing cycle (IPRO-1C).]{
\includegraphics[width=\textwidth]{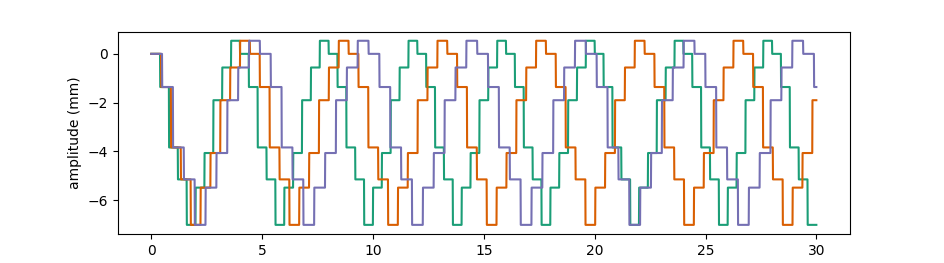}
}
\caption{Amplitudes corresponding to the first 30 seconds of three scenarios for patient 2.}
\label{scenarios}
\end{figure}

We generated each evaluation motion scenario by randomly concatenating breathing cycles from the evaluation set. For each beam, the motion state sequence was constructed by drawing (with uniform probability and with replacement) and concatenating breathing cycles until the total breathing time corresponding to the sequence was sufficient to cover the duration of the beam delivery. Due to the variation in period between the breathing cycles, this method implicitly generates phase variation within the set of generated breathing patterns and allows for studying the interplay effect under various motion scenarios. It implicitly relies on the assumption that there is no uncertainty about the breathing cycle phase at the start of the delivery of each beam, as each breathing cycle begins in the end-exhale phase. Figure \ref{shuffle_scenarios} shows three scenarios generated using the described method.

\subsection{Treatment evaluation}

We evaluated the robustness of each investigated treatment plan by simulating the plan delivery -- without gating or re-scanning -- in multiple motion scenarios, generated as described in Section \ref{scenario_modeling}. The time structure of the delivery was as described in Section \ref{time_structure}. Combining a motion scenario, a treatment plan, and its corresponding time structure enabled 4DDC with the methodology described in Section \ref{4DDC}. The number of evaluation motion scenarios per plan was 500, as recommended by Pastor-Serrano et al.\ for statistically accurate analysis of dose uncertainties in the presence of interplay \cite{pastor-serrano_how_2021}. Consequently, we analyzed 500 dose distributions per investigated plan by their variation of various dosimetric qualities and the objective function value.

\subsection{Treatment planning}

Treatment planning was performed using a RayStation (RaySearch Laboratories AB, Stockholm, Sweden) machine model of an IBA system with the Dedicated PBS nozzle. All compared plans originated from the same initial \textit{template plan}. This plan included three beams with angles chosen based on the patient's geometry. It was optimized with 40 iterations of 4DRO \cite{janson_treatment_2024}, with the scenarios being the motion states from a full breathing cycle containing the reference CT. Setup and range uncertainties were not considered. The template plan acted as a basis for all the compared plans in three ways: First, it distributed sufficiently many spots around the CTV to account for the full range of motion and ensured that all compared plans used the same set of spots. Second, it allowed the selection of objective functions and weights that complied with the treatment's dosimetric goals. Third, it provided an improved and consistent initial point for the numerical algorithm employed in the optimization of the compared plans.

The objective functions used in the optimization of all plans, including the template plan, were designed per patient to comply with the dosimetric criteria from RTOG 1308 \cite{giaddui_establishing_2016}, but limited to the available OAR segmentations, and scaled to a single fraction with 200 cGy as the prescribed dose. To obtain target coverage, we increased the objective weight of the CTV from an initial guess until the CTV D99 reached the prescribed dose in all motion states or until the weight increases did not result in any further improvements (which occurred for one CT when the CTV D99 was 99.5\% of the prescribed dose). In the case where OAR objectives were used, they were designed to penalize dose above the dose levels specified in the criteria from RTOG 1308. Due to a mismatch between the segmented OARs and those listed in RTOG 1308, an objective was used for the medulla instead of the spinal cord for CT 1. A complete list of objectives used for each CT geometry can be found in Appendix \ref{optimization_functions}.

\subsection{Optimization methods}

All optimization methods employed 40 iterations of sequential quadratic programming (SQP), with the initial point as the spot weights from the template plan and without any further spot weight filtering.

\begin{itemize}
    \item \textbf{4D Robust (4DRO)}: This plan was generated by 40 additional iterations of the SQP solver in RayStation and acted as a reference for the results of the interplay-driven plans. 
\end{itemize}

\noindent Except for 4DRO, the investigated plans were all optimized using IPRO and differ only in the scenario set $\mathcal{S}$, under different assumptions on the availability of representative breathing data. They were optimized using 40 iterations of the SQP solver SNOPT (7.7 Stanford Business Software, Stanford, California), interfaced with Matlab (2024a, Mathworks Inc, Natick, Massachusetts, USA). To smoothen the objective in Problem \eqref{IPRO}, the max-operator was approximated with the weighted-power-mean operator with the power parameter set to 8.

\begin{itemize}
    \item \textbf{Non-robust interplay-driven optimization (IPO)}: To study the performance of plans optimized considering a single breathing cycle, we implemented IPO with the breathing scenario taken as the repetition of the first breathing cycle available in the data.

    \item \textbf{Interplay-robust optimization with repeated breathing (IPRO-1C)}: Because it may be infeasible to acquire motion states from multiple breathing cycles for treatment planning, this first IPRO method used a scenario set consisting only of scenarios which repetitions of the first breathing cycle, but shifted in start phase and scaled in breathing period. Three sets of scenarios were investigated: one using breathing period scaling factors in $\{ 0.9, 1, 1.1 \}$ and start phase shifts in $\{ -1, 0, +1 \}$ (9 scenarios), the second using breathing period scaling factors in $\{ 0.8, 0.9, 1, 1.1, 1.2 \}$ and start phase shifts in $\{-2, -1, 0, +1, +2 \}$ (25 scenarios), and the third with breathing period scaling factors in $\{ 0.7, 0.8, 0.9, 1, 1.1, 1.2, 1.3 \}$ and start phase shifts in $\{-3, -2, -1, 0, +1, +2, +3 \}$ (49 scenarios).
    
    \item \textbf{Interplay-robust optimization with simulated breathing (IPRO)}: To study the extent to which further robustness gains can be made by including additional breathing cycles in the optimization, we considered a scenario set generated by the same method as the set of evaluation scenarios but by randomly selecting breathing cycles for concatenation from the optimization set. This method was investigated for 9, 25, and 49 scenarios.
    
\end{itemize}

\noindent A challenge of IP(R)O(-1C) is that the delivery time structure changes with the variable spot weights. As in previous literature \cite{bernatowicz_advanced_2017, engwall_4d_2018}, we address this heuristically by updating the delivery time structure according to the current spot weights at regular intervals (every 10 iterations) during the optimization process.

\section{Results}

Here, we present the performance of each planning method in each of the investigated patient cases. Performance is measured by considering the variation in dosimetric criteria over the evaluated motion scenarios. For the different cases, we show in Figures \ref{patient0_CTV_D98}--\ref{patient1_v2_lung_mean_dose_alt} boxplots of the CTV D98, the CTV homogeneity index (HI = $\frac{\text{D5 - D95}}{\text{D50}}$), and dose to a relevant OAR. Each figure also shows the variation of the considered metric between the phase doses used in the optimization of 4DRO (each deformed to the planning image). These \textit{static} doses are unrealistically good for two reasons: they are computed without considering the interplay effect, and they are evaluated on the same images as they were optimized (unlike the evaluated methods). Still, they indicate the ideal performance and are included for reference.

Because of the different implicit trade-offs made in the optimization between conflicting objectives, it is difficult to determine the most favorable method from the unprocessed results. To address this issue, we normalized all doses to have equal near-worst-case target coverage. More precisely, all evaluation doses were scaled by a factor to make the 5th percentile D98 equal to that of the evaluation doses for the 4DRO, acting as the benchmark. For simplicity, the effect on the time structure was ignored. The 95th percentile OAR doses, as well as the CTV D2, are then shown in Table \ref{oar_doses}. One way to compare the methods without having to consider trade-offs between conflicting criteria is to compare the impact on the objective function used in optimization. Therefore, we show the variations over the evaluation scenarios of the objective value in Appendix \ref{objective_values}.

\begin{table}
    \centering
    \makebox[\textwidth]{%
    \begin{tabular}{c|c|c|>{\columncolor{cyan!10}}p{1cm}|c|p{1cm}|p{1cm}|p{1.2cm}|p{1cm}|p{1cm}|p{1cm}}
         Case &  ROI &   Metric &4DRO (cGy) &  IPO &  IPRO-1C-9S &  IPRO-1C-25S &  IPRO-1C-49S &  IPRO-9S &  IPRO-25S & IPRO-49S \\
         \hline
        \multicolumn{4}{c}{ } & \multicolumn{7}{|c}{Change (\%) relative to 4DRO} \\
        \hline
         \multirow{7}{*}{1} &  \multirow{1}{*}{CTV} &   D2 &223.4&  \textcolor{red}{+3.8} &  +0.1&  +0.1&  -0.4&  +0.4&  -0.8& \textcolor{blue}{\textbf{-1.7}} \\
         \cline{2-11}&  \multirow{2}{*}{Heart} &   D2 &61.3&  +0.9&  \textcolor{blue}{-1.9} &  \textcolor{blue}{-2.5} &  \textcolor{blue}{-2.4} &  \textcolor{blue}{-1.7} &  \textcolor{blue}{-3.2} & \textcolor{blue}{\textbf{-4.2}} \\
         \cline{3-11}&  &   Mean &3.8&  +0.9&  \textcolor{blue}{-1.6} &  \textcolor{blue}{-2.0} &  \textcolor{blue}{-1.9} &  \textcolor{blue}{-1.6} &  \textcolor{blue}{-2.8} & \textcolor{blue}{\textbf{-3.7}} \\
         \cline{2-11}&  \multirow{2}{*}{Medulla} &   D2 &86.1&  \textcolor{red}{+1.9} &  \textcolor{blue}{-2.5} &  \textcolor{blue}{-1.7} &  \textcolor{blue}{-2.3} &  +0.7&  \textcolor{blue}{-3.0} & \textcolor{blue}{\textbf{-3.2}} \\
         \cline{3-11}&  &   Mean &7.7&  +0.7&  \textcolor{blue}{-2.9} &  \textcolor{blue}{-2.8} &  \textcolor{blue}{-3.1} &  \textcolor{blue}{-1.5} &  \textcolor{blue}{-3.9} & \textcolor{blue}{\textbf{-4.2}} \\
         \cline{2-11}&  \multirow{2}{*}{Esophagus} &   D2 &202.5&  \textcolor{red}{+2.0} &  -0.7&  \textcolor{blue}{-2.5} &  \textcolor{blue}{-2.3} &  \textcolor{blue}{-1.9} &  \textcolor{blue}{-2.6} & \textcolor{blue}{\textbf{-3.0}} \\
         \cline{3-11}&  &   Mean &32.7&  \textcolor{red}{+2.7} &  +0.4&  \textcolor{blue}{-1.4} &  \textcolor{blue}{-1.0} &  -0.9&  \textcolor{blue}{-1.4} & \textcolor{blue}{\textbf{-2.0}} \\
         \cline{2-11}&  \multirow{2}{*}{Normal lung} &   D2 &206.8&  \textcolor{red}{+1.9} &  \textcolor{blue}{-1.1} &  \textcolor{blue}{-1.5} &  \textcolor{blue}{-1.7} &  -0.9&  \textcolor{blue}{-2.0} & \textcolor{blue}{\textbf{-2.7}} \\
 \cline{3-11}& & Mean & 30.5& +0.4& \textcolor{blue}{-2.5} & \textcolor{blue}{-2.9} & \textcolor{blue}{-3.2} & \textcolor{blue}{-2.4} & \textcolor{blue}{-3.5} &\textcolor{blue}{\textbf{-4.3}} \\
 \hline
 \hline
 \multirow{5}{*}{2} & \multirow{1}{*}{CTV} & D2 & 222.7& \textcolor{red}{+1.9} & -0.2& \textbf{-0.9} & \textbf{-0.9} & -0.8& -0.7&-0.6 \\
 \cline{2-11}& \multirow{2}{*}{Heart} & D2 & 172.2& \textcolor{blue}{-2.6} & \textcolor{blue}{-3.4} & \textcolor{blue}{-4.0} & \textcolor{blue}{\textbf{-5.3}} & \textcolor{blue}{-4.7} & \textcolor{blue}{-5.1} &\textcolor{blue}{-4.6} \\
 \cline{3-11}& & Mean & 11.4& -0.1& -0.9& \textcolor{blue}{-2.2} & \textcolor{blue}{\textbf{-2.9}} & \textcolor{blue}{-2.1} & \textcolor{blue}{-2.5} &\textcolor{blue}{-1.4} \\
 \cline{2-11}& \multirow{2}{*}{Normal lung} & D2 & 164.2& -0.1& \textcolor{blue}{-2.6} & \textcolor{blue}{\textbf{-3.3}} & \textcolor{blue}{-3.2} & \textcolor{blue}{-2.9} & \textcolor{blue}{-2.9} &\textcolor{blue}{-3.2} \\
 \cline{3-11}& & Mean & 21.6& -0.6& \textcolor{blue}{-2.8} & \textcolor{blue}{-3.4} & \textcolor{blue}{\textbf{-3.6}} & \textcolor{blue}{-3.2} & \textcolor{blue}{-3.2} &\textcolor{blue}{-3.4} \\
 \hline
        \hline
 \multirow{3}{*}{3a} & \multirow{1}{*}{CTV} & D2 & 230.2& +0.5& -0.2& \textcolor{blue}{-2.1} & \textcolor{blue}{-1.6} & \textcolor{blue}{-1.6} & \textcolor{blue}{\textbf{-3.1}} &\textcolor{blue}{-2.6} \\
 \cline{2-11}& \multirow{2}{*}{Normal lung} & D2 & 188.7& +0.8& -0.5& \textcolor{blue}{-1.9} & \textcolor{blue}{-2.8} & \textcolor{blue}{-1.2} & \textcolor{blue}{\textbf{-4.5}} &\textcolor{blue}{\textbf{-4.5}} \\
 \cline{3-11}& & Mean & 13.7& -0.3& \textcolor{blue}{-1.5} & \textcolor{blue}{-2.7} & \textcolor{blue}{-3.8} & \textcolor{blue}{-1.8} & \textcolor{blue}{-5.2} &\textcolor{blue}{\textbf{-5.5}} \\
 \hline
        \hline
 \multirow{3}{*}{3b} & \multirow{1}{*}{CTV} & D2 & 238& \textcolor{red}{+1.6} & +0.7& -0.2& \textcolor{blue}{-1.8} & \textcolor{blue}{-1.0} & \textcolor{blue}{-3.4} &\textcolor{blue}{\textbf{-3.6}} \\
 \cline{2-11}& \multirow{2}{*}{Normal lung} & D2 & 185.2& -0.4& \textcolor{blue}{-1.1} & \textcolor{blue}{-2.1} & \textcolor{blue}{-3.6} & \textcolor{blue}{-3.5} & \textcolor{blue}{-7.5} &\textcolor{blue}{\textbf{-8.0}} \\
 \cline{3-11}& & Mean & 14& -0.9& \textcolor{blue}{-2.1} & \textcolor{blue}{-3.3} & \textcolor{blue}{-4.8} & \textcolor{blue}{-4.9} & \textcolor{blue}{-9.1} &\textcolor{blue}{\textbf{-9.7}} \\
 \hline
        \hline
 \multirow{3}{*}{3b*} & \multirow{1}{*}{CTV} & D2 & - & - & +0.5& \textcolor{red}{+1.1} & \textcolor{red}{+2.5} & - & - &- \\
 \cline{2-11}& \multirow{2}{*}{Normal lung} & D2 & - & - & \textcolor{blue}{-4.5} & \textcolor{blue}{-4.7} & \textcolor{blue}{-4.7} & - & - &- \\
 \cline{3-11}& & Mean & - & - & \textcolor{blue}{-5.2} & \textcolor{blue}{-5.7} & \textcolor{blue}{-6.2} & - & - &- \\
    \end{tabular}
    }
    \caption{The patient-specific dosimetric results after normalizing the doses for each method to match the 5th percentile CTV D98 for 4DRO. \textcolor{red}{Red} and \textcolor{blue}{blue} values indicate an increase or decrease by $1\%$ or more with respect to 4DRO, respectively. The prescribed dose was 200 cGy. The \textbf{bold} values indicate the best-performing method with respect to the relevant metric. The asterisk indicates the version of case 3b with a different breathing cycle used for IPRO-1C.}
    \label{oar_doses}
\end{table}

\subsection{Case 1 - Large CTV}

CT 1 had the largest CTV, resulting in plans with 105 energy layers and 12716 spots. The dosimetric performance is shown in Figures \ref{patient0_CTV_D98}--\ref{patient0_medulla_mean_dose}, where Subfigure \ref{patient0_medulla_mean_dose} shows the mean dose to the medulla, the OAR contributing the most to the objective function (Table \ref{objectives}).

Compared to the plan made with 4DRO, acting as the benchmark, the plans optimized with IPRO(-1C) consistently increased the CTV D98 in both the mean and 5th percentile. This was achieved without a clear increase to the CTV HI or the medulla mean dose. In contrast, IPO failed to match the CTV D98 or the CTV HI of 4DRO. Considering the effect of IPRO(-1C) after normalization, the CTV D2 was not drastically affected, with changes of less than $\pm 1\%$. The effect on OAR dose was more apparent, with typical decreases by more than $1\%$ for most IPRO(-1C) methods, and IPRO-49S consistently performing the best with decreases in mean dose by $2.0 - 4.3 \%$ and in D2 by $2.7 - 4.2 \%$.

\begin{figure}
\centering
\makebox[\textwidth][c]{
\subfloat[Case 1: CTV D98]{
\includegraphics[width=0.33\textwidth]{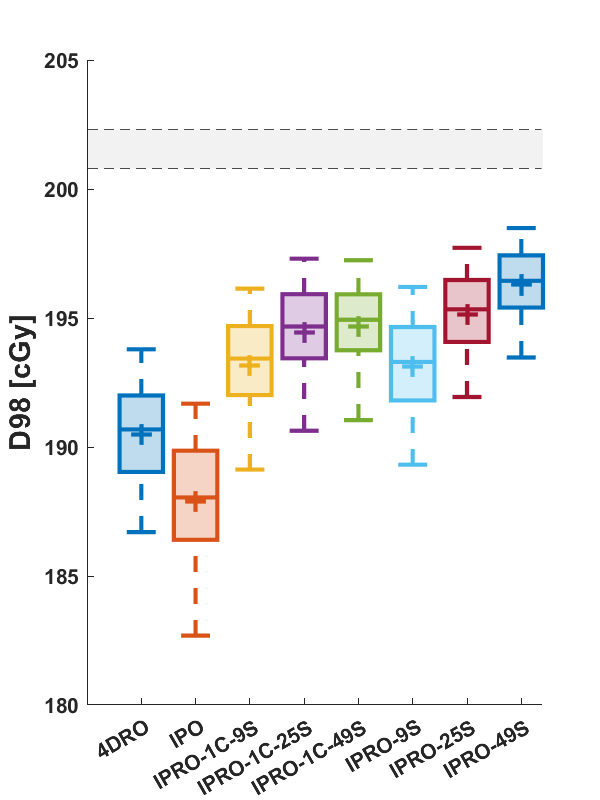}
\label{patient0_CTV_D98}
}
\subfloat[Case 1: CTV HI]{
\includegraphics[width=0.33\textwidth]{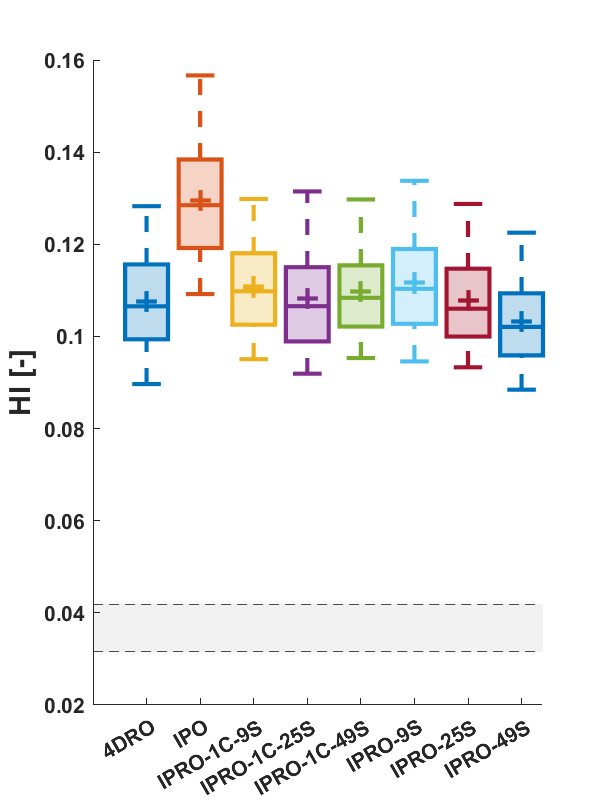}
\label{patient0_CTV_HI}
}
\subfloat[Case 1: Medulla mean dose]{
\includegraphics[width=0.33\textwidth]{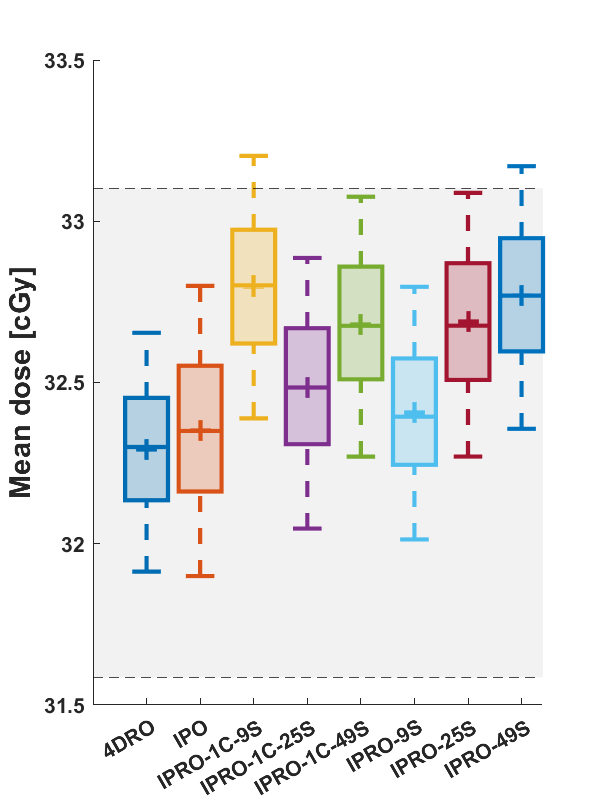}
\label{patient0_medulla_mean_dose}
}}

\makebox[\textwidth][c]{
\subfloat[Case 2: CTV D98]{
\includegraphics[width=0.33\textwidth]{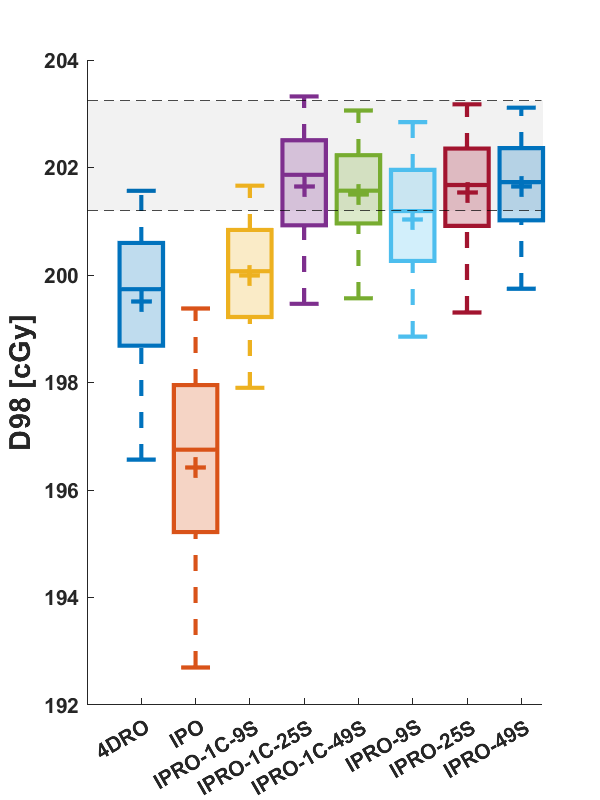}
\label{patient2_CTV_D98}
}
\subfloat[Case 2: CTV HI]{
\includegraphics[width=0.33\textwidth]{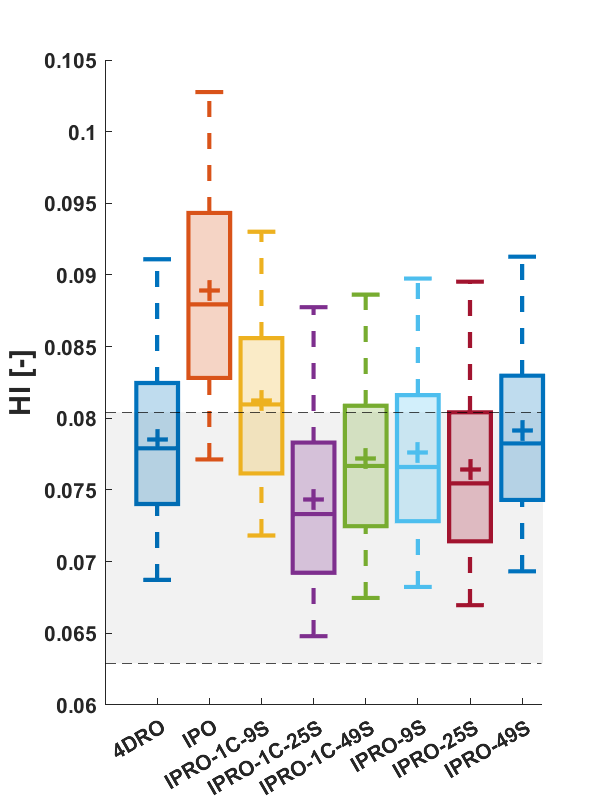}
\label{patient2_CTV_HI}
}
\subfloat[Case 2: Heart mean dose]{
\includegraphics[width=0.33\textwidth]{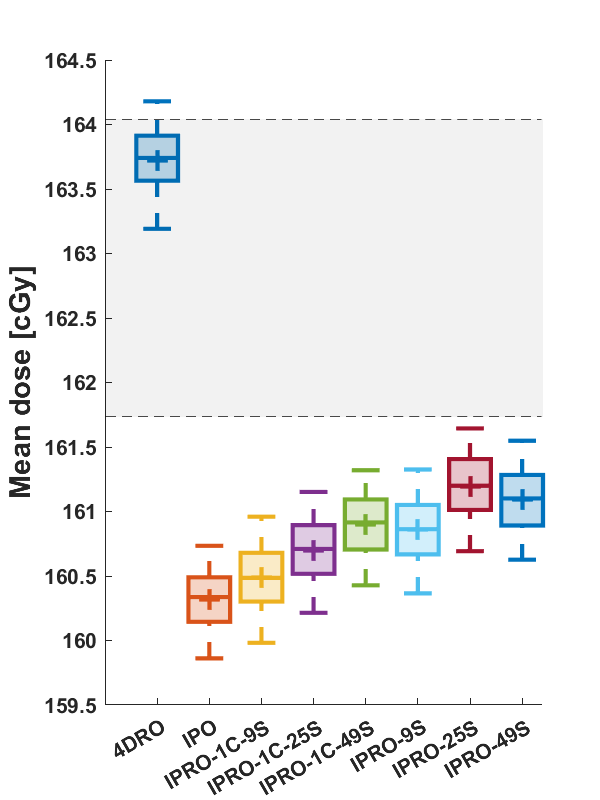}
\label{patient2_heart_mean_dose}
}}
\caption{Distributions of CTV and medulla dose statistics for each of the investigated optimization methods for cases 1 and 2. The boxes indicate the inter-quartile range, while the whiskers indicate the 5th and 95th percentiles. For each box, the distribution mean and median are indicated by the plus and the solid line, respectively. The box edges indicate the 25th and 75th percentiles, and the whiskers indicate the 5th and 95th percentiles. Additionally, the shaded area specifies the spread of values among the optimized phase doses achieved by 4DRO (deformed to the planning image) and acts as an indication of the desired dosimetric trade-offs.}
\label{boxplots}
\end{figure}

\subsection{Case 2 - Smaller motion amplitude}

Case 2 required 80 energy layers and 6834 spots to cover the CTV. The motion amplitude was approximately half that of the other cases, suggesting less severe consequences of the interplay effect. This was confirmed by the CTV D98, which was considerably higher than for the other patient cases, as shown in Figures \ref{patient2_CTV_D98}--\ref{patient2_heart_mean_dose}. For example, the mean and median of IPRO(-1C) with 25 or more scenarios were inside the interval based on the 4DRO static doses.

Considering the normalized doses (Table \ref{oar_doses}), all the interplay-based methods decreased the heart D2 by at least $2.6\%$. For the robust methods, the decrease was $ 3.4 - 5.3 \%$, with the greatest decrease from IPRO-1C-49S. For normal lung, IPRO(-1C) decreased D2 (mean) by $2.6 - 3.3 \%$ ($2.8 - 3.6 \%$), while the corresponding decrease for IPRO was $2.9 - 3.2 \%$ ($3.2 - 3.4 \%$).

\subsection{Case 3a - Small CTV}

CT 3 had the smallest CTV, requiring only 54 energy layers and 3439 spots distributed across the three beams. The size of the CTV made treatment planning more difficult, resulting in greater dose heterogeneity for all plans. As optimization objectives were applied only to the CTV, the mean dose to the normal lung (both lungs minus CTV) is shown alongside the CTV statistics in Figures \ref{patient1_CTV_D98}--\ref{patient1_lung_mean_dose} to indicate the dosimetric effect outside the CTV.

IPRO(-1C) consistently increased the mean and near-worst-case for both CTV D98 and CTV HI compared to 4DRO. After normalizing the doses, IPRO with 25 and 49 scenarios performed similarly (Table \ref{oar_doses}), improving CTV D2 by $2.6 - 3.1 \%$ and normal lung D2 and mean dose by $4.5 \%$ and $5.2 - 5.5 \%$, respectively.

\begin{figure}[]
\ContinuedFloat
\centering
\makebox[\textwidth][c]{
\subfloat[Case 3a: CTV D98]{
\includegraphics[width=0.33\textwidth]{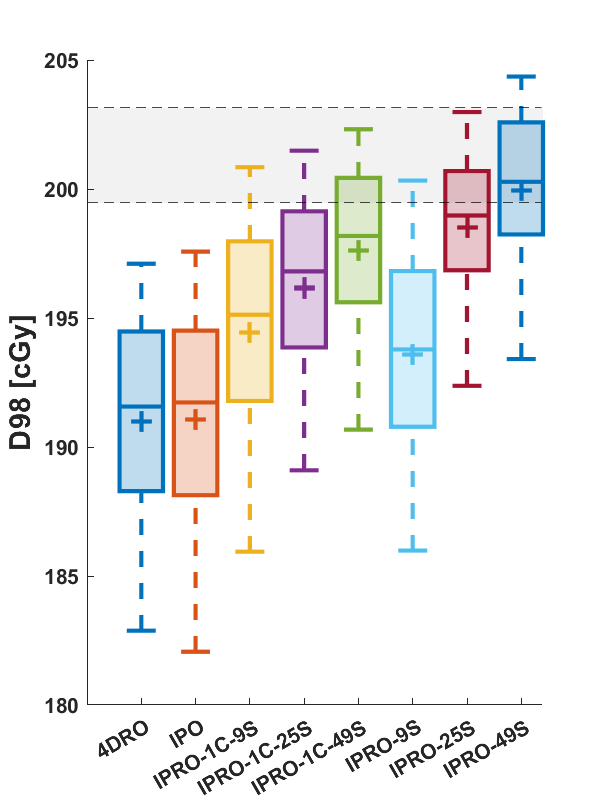}
\label{patient1_CTV_D98}
}
\subfloat[Case 3a: CTV HI]{
\includegraphics[width=0.33\textwidth]{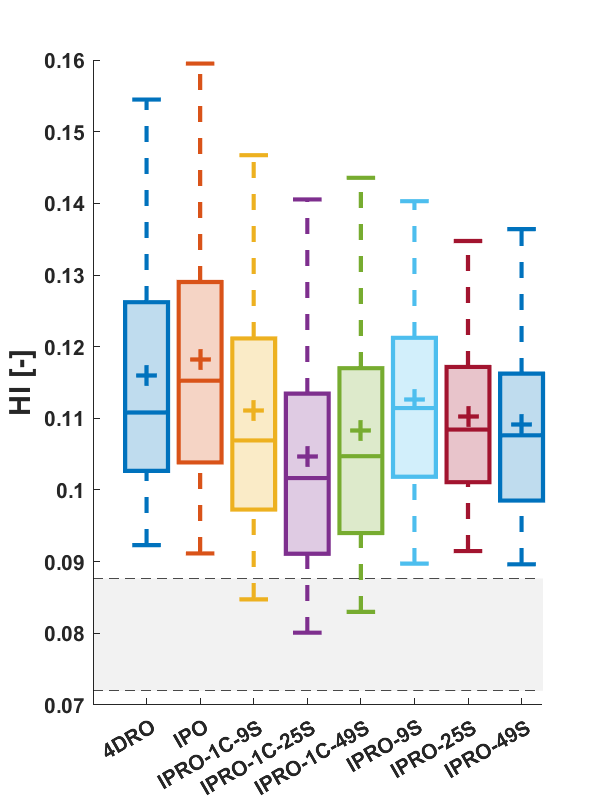}
\label{patient1_CTV_HI}
}
\subfloat[Case 3a: Normal lung mean dose]{
\includegraphics[width=0.33\textwidth]{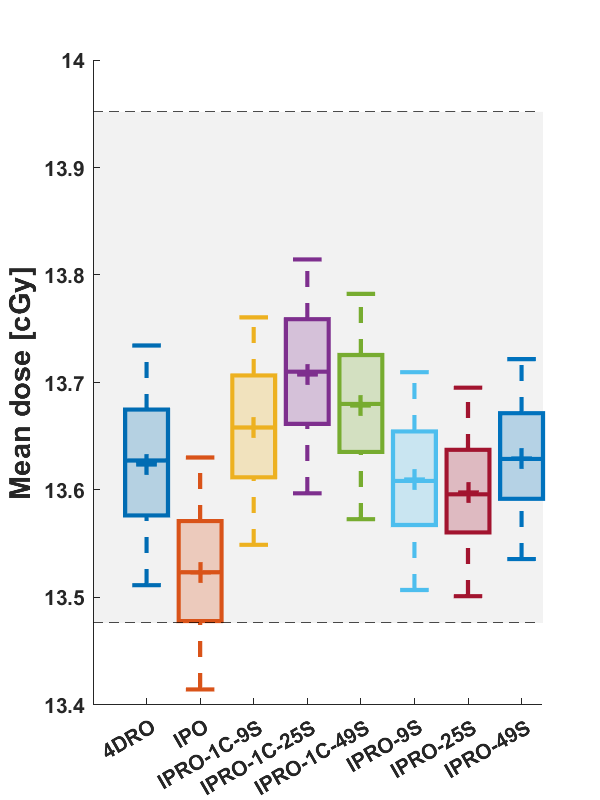}
\label{patient1_lung_mean_dose}
}}
\makebox[\textwidth][c]{
\subfloat[Case 3b: CTV D98]{
\includegraphics[width=0.33\textwidth]{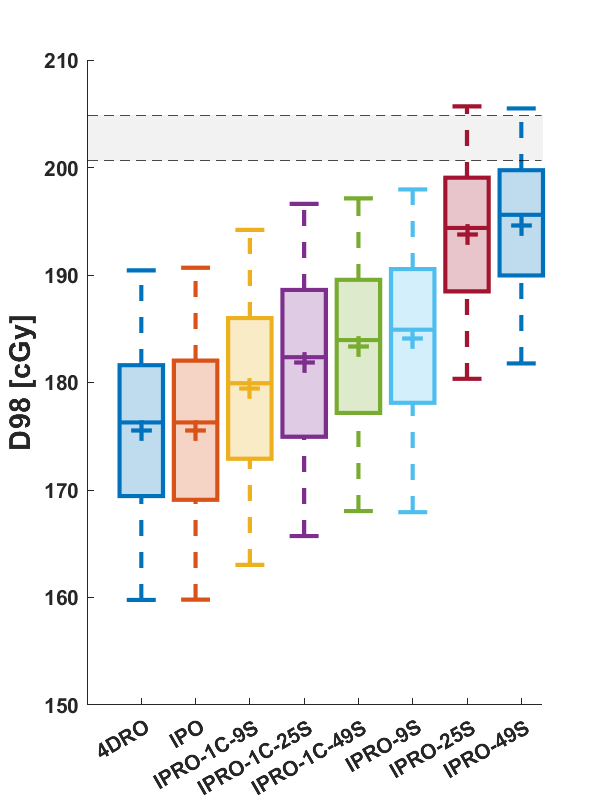}
\label{patient1_v2_CTV_D98}
}
\subfloat[Case 3b: CTV HI]{
\includegraphics[width=0.33\textwidth]{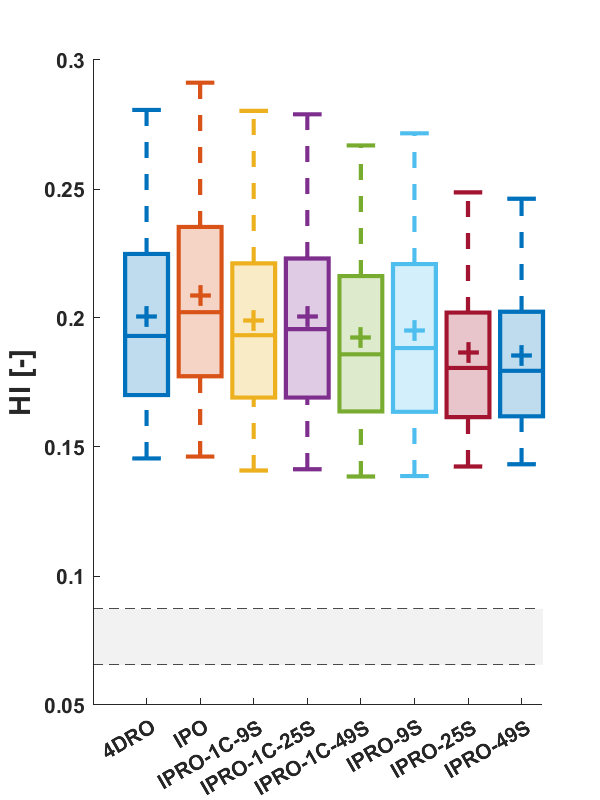}
\label{patient1_v2_CTV_HI}
}
\subfloat[Case 3b: Normal lung mean dose]{
\includegraphics[width=0.33\textwidth]{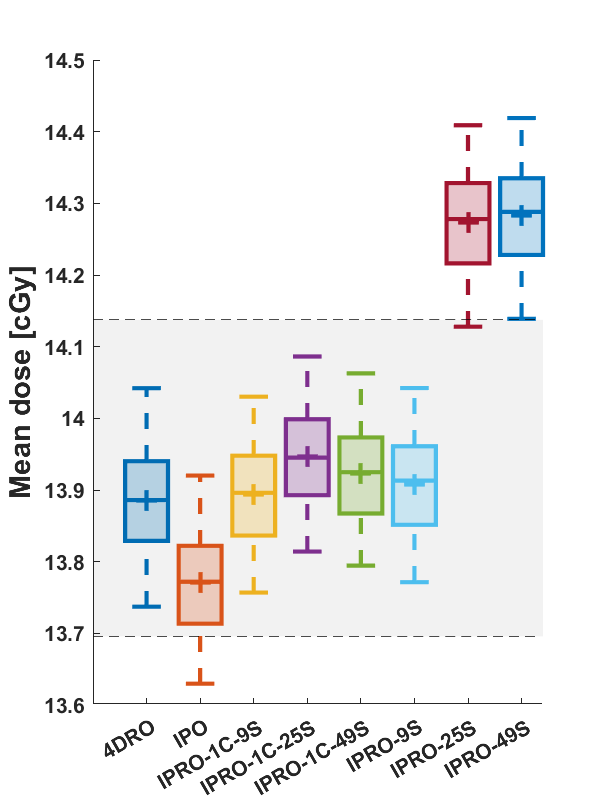}
\label{patient1_v2_lung_mean_dose}
}}
\makebox[\textwidth][c]{
\subfloat[Case 3b*: CTV D98]{
\includegraphics[width=0.33\textwidth]{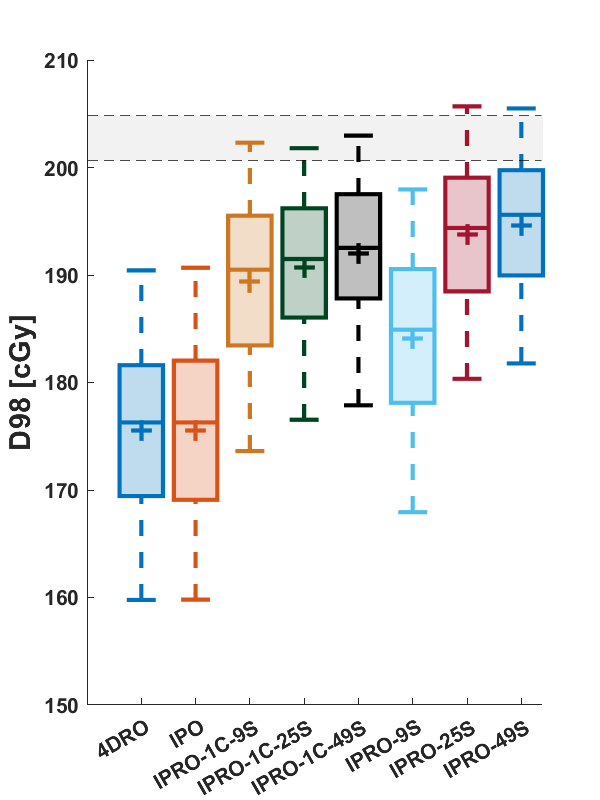}
\label{patient1_v2_CTV_D98_alt}
}
\subfloat[Case 3b*: CTV HI]{
\includegraphics[width=0.33\textwidth]{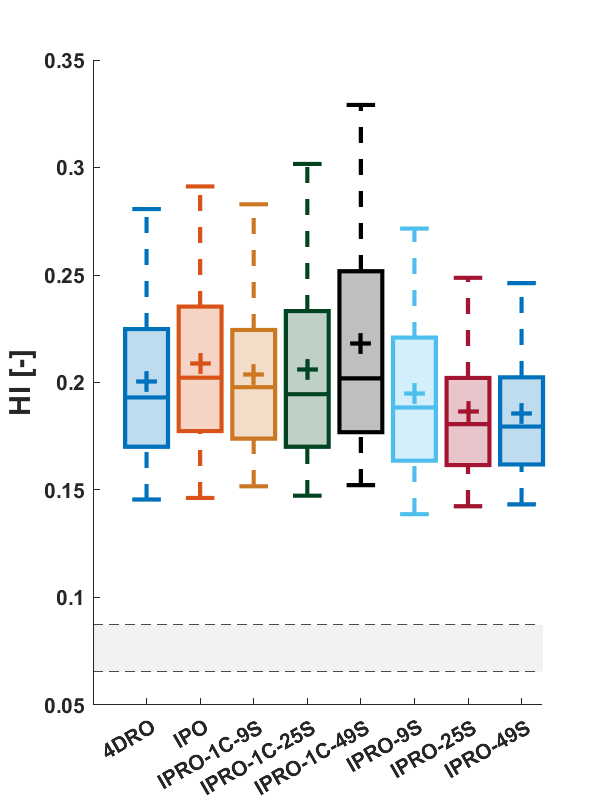}
\label{patient1_v2_CTV_HI_alt}
}
\subfloat[Case 3b*: Normal lung mean dose]{
\includegraphics[width=0.33\textwidth]{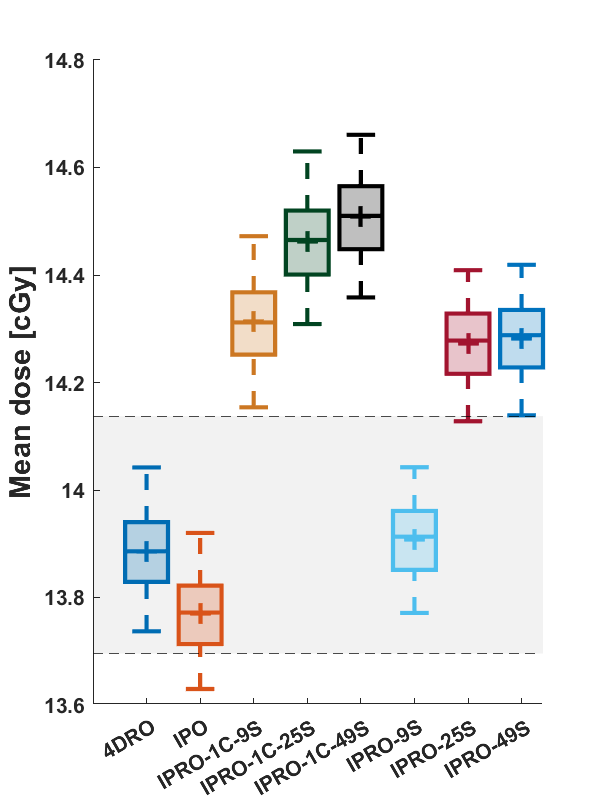}
\label{patient1_v2_lung_mean_dose_alt}
}}
\caption{(continued) Distributions of CTV and OAR dose statistics for each of the investigated optimization methods for cases 3a, 3b, and 3b* (in which the alternative breathing cycle was used for IPRO-1C).}
\end{figure}

\subsection{Case 3b - Second motion pattern}

The results for case 3b, with the second motion pattern, are shown in Figures \ref{patient1_v2_CTV_D98}--\ref{patient1_v2_lung_mean_dose}. As seen in Figure \ref{motion}, the second motion pattern differs from the first in that the first breathing cycle has much less amplitude variation than the maximal of all cycles within the motion pattern. Consequently, the difference between IPRO-49S, which performed the best in CTV D98 and CTV HI, and IPRO-1C-49S was larger than for the other cases. Therefore, we repeated the experiment, calling it 3b*, with another breathing cycle to generate the scenario set for IPRO-1C. The consequence, shown in Figures \ref{patient1_v2_CTV_D98_alt}--\ref{patient1_v2_lung_mean_dose_alt}, was a clear increase in CTV D98 and normal lung mean dose.

The normalization of target doses (Table \ref{oar_doses}) shows that IPRO-49S performed the best, even after the adjustments to the planning breathing cycle used for IPRO-1C, reducing the CTV D2, the normal lung D2, and the normal lung mean dose by $3.6\%$, $8.0\%$, and $9.7\%$, respectively. However, the effect of the change of breathing cycle was apparent regardless of the number of scenarios. In particular, IPRO-1C-9S improved in every metric and achieved results comparable to IPRO-9S.

\begin{figure}[H]
\centering

\end{figure}

\begin{figure}[H]
\centering

\end{figure}

\section{Discussion}

We implemented interplay-robust optimization for patient cases with synthetically generated irregular breathing motion. The results show that IPRO consistently improves both mean and near-worst-case CTV D98, with little effect on CTV dose homogeneity, for PBS deliveries in the investigated cases. In most cases, these results are achieved without compromising OAR-sparing. In addition, the OAR doses -- whose functions were typically of low weight in the optimization objective  -- typically varied little compared to their variation within the static doses used in 4DRO. In some cases where IPRO(-1C) substantially improved the target coverage, the integral dose (measured by the mean dose in the normal lung and not included in the optimization) increased to be outside of the interval spanned by the static doses. However, when normalizing all methods to equal 4DRO in target coverage, the OAR-sparing effects of IPRO(-1C) were apparent (Table \ref{oar_doses}). This result is most apparent for IPRO, based on optimization scenarios sampled using the same technique used to sample evaluation scenarios. The result also holds -- although with smaller improvements -- for IPRO-1C, based on using a single breathing cycle to generate the optimization scenarios deterministically.

Between the scenario generation methods for IPRO(-1C), there was typically a positive impact from increasing the number of scenarios, with few exceptions. Likewise, using multiple breathing cycles to generate the optimization scenarios (IPRO) was advantageous compared to using a single cycle. This difference was most apparent for case 3b, for which the improvement in 5th percentile CTV D98 with IPRO-1C was less than half of that from using IPRO, which showed that IPRO-1C with a single breathing cycle unrepresentative of the evaluation motion scenarios may fail to generate robust plans. However, we then showed (with Case 3b*) that this discrepancy can be decreased by using a breathing cycle with sufficient amplitude range (Figures \ref{patient1_v2_CTV_D98_alt}--\ref{patient1_v2_lung_mean_dose_alt}).

A limitation of this work is the disregard of ribcage motion in the synthetization of motion states. This follows from the method used to generate s4DCT in Duetschler et al.\ \cite{duetschler_synthetic_2022}. Consequently, although our study has generalized the results from Engwall et al.\ \cite{engwall_4d_2018} to irregular motion, further work that investigates IPRO with an even more realistic representation of patient breathing motion is desirable. We believe that the improvement presented by IPRO over 4DRO would hold regardless of the exact patient breathing model.

Furthermore, no range or setup errors were considered, and the impact of real-time motion management techniques such as gating or re-scanning was not investigated. Although we believe that there would still be an advantage of IPRO due to the more realistic dose delivery model, it is possible that its advantage would decrease in magnitude when other uncertainties are considered, or when real-time motion management techniques are applied. An interesting future direction of research would be to investigate to which extent the number of re-scans could be reduced with the use of IPRO, without compromising target coverage or dose homogeneity.

Instead, the case that may compromise the robustness aimed at by IPRO is when the motion states that may occur during treatment delivery are not represented in the scenarios used during optimization. This was the case with the initial results for IPRO-1C for case 3b and it is possibly also the reason for the performance difference between IPRO with 9, 25, and 49 scenarios. In this study, we selected the optimization and evaluation sets to be unique but statistically similar (Table \ref{breathing_statistics}). Larger discrepancies between the data used for optimization and evaluation, respectively, are expected to decrease the robustness gain of IPRO. Non-robust IPO was included in the numerical experiments primarily to highlight this effect; the accuracy of PBS allows highly conformal plans that are very sensitive to delivery uncertainty. To further improve IPRO-based treatment planning, work is needed to establish the most appropriate representation of the patient's motion, including uncertainties. The computational demands of representing a large set of breathing scenarios and the associated dose influence matrices posed a challenge in this study. The development of faster and more memory-efficient methods for dose computation, motion modeling, and optimization will help to accurately represent, evaluate, and mitigate patient motion on current hardware.

Lastly, we want to emphasize the fact that IPRO-1C-9S is not a more computationally intensive task than 4DRO; it uses the same number of motion states and performs the same number of dose computations. Instead of contour sets on each of the included motion states, deformable registrations are needed to accumulate the dose on the reference state. The principal difference is the more accurate model of how the dose will be delivered during treatment, explicitly considering time structures. The unaffected computation effort is important, as our results have indicated that IPRO-1C-9S leads to dosimetric improvements over 4DRO. Although no implementation of 4D optimization may be sufficient to fully replace the use of other motion management techniques, further robustness gains are likely to alleviate some of the need for either re-scanning, breath hold, or gating, leading toward treatments of shorter duration and increased patient comfort.

\section{Conclusion}

In this study, we implemented different variants of interplay-robust optimization and evaluated the resulting plans based on synthetically generated 4DCTs exhibiting irregular breathing. The findings indicate the advantages of using IPRO over conventional 4DRO. Including more scenarios and breathing cycles in the optimization typically led to the largest dosimetric improvements, highlighting the importance of accurately representing the breathing variation. However, the explicit consideration of the interplay effect by generating breathing scenarios using a single breathing cycle (IPRO-1C) already presents improvements over 4DRO.

\appendix

\section{Optimization objective functions} \label{optimization_functions}

The geometry-specific objective functions are presented in Table \ref{objectives}.

\begin{table}[]
    \centering
    \begin{tabular}{c|c|c|c|c}
        \hline
        Geometry & ROI & Function type & Dose level (cGy) & Weight \\
        \hline
        \multirow{3}{*}{CT 1} & CTV & Min. dose & 201 & 400 \\
        \cline{2-5}
         & CTV + 1cm & Max. dose & 210 & 50 \\
        \cline{2-5}
         & Heart & Max. dose & 86 & 1 \\
         \cline{2-5}
         & Heart & Max dose & 129 & 1 \\
         \cline{2-5}
         & Medulla & Max. dose & 143 & 100 \\
         \cline{2-5}
         & Esophagus & Max. dose & 211 & 1 \\
        \hline
        \multirow{2}{*}{CT 2} & CTV & Min. dose & 201 & 1200 \\
        \cline{2-5}
         & CTV + 1 cm & Max. dose & 210 & 10 \\
         \cline{2-5}
         & Heart & Max. dose & 86 & 1 \\
         \cline{2-5}
         & Heart & Max. dose & 129 & 1 \\
        \hline
        \multirow{2}{*}{CT 3} & CTV & Min. dose & 201 & 80 \\
        \cline{2-5}
         & CTV + 1cm & Max. dose & 210 & 1 \\
        \hline
    \end{tabular}
    \caption{The patient-specific objectives used to optimize plans for all methods.}
    \label{objectives}
\end{table}

\section{Objective values} \label{objective_values}

\begin{figure}[H]
\centering
\makebox[\textwidth][c]{
\subfloat[Case 1]{
\includegraphics[width=0.4\textwidth]{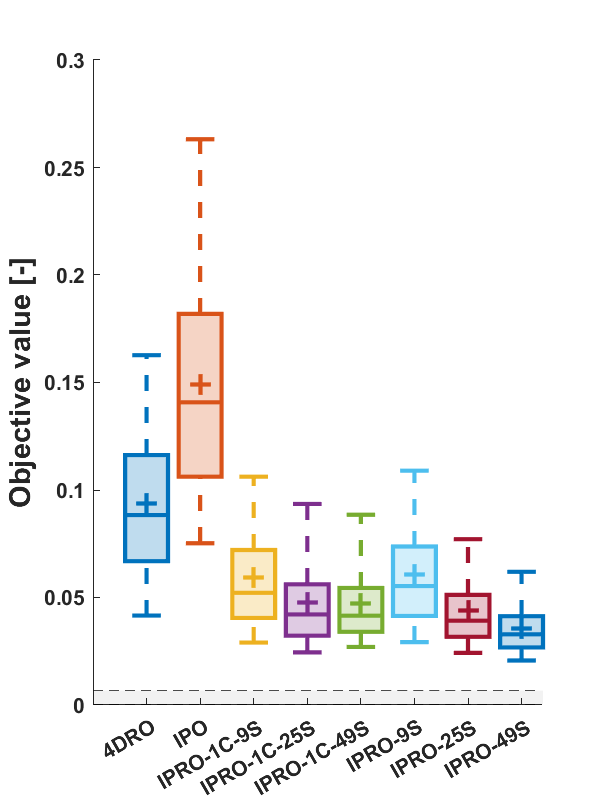}
}
\subfloat[Case 2]{
\includegraphics[width=0.4\textwidth]{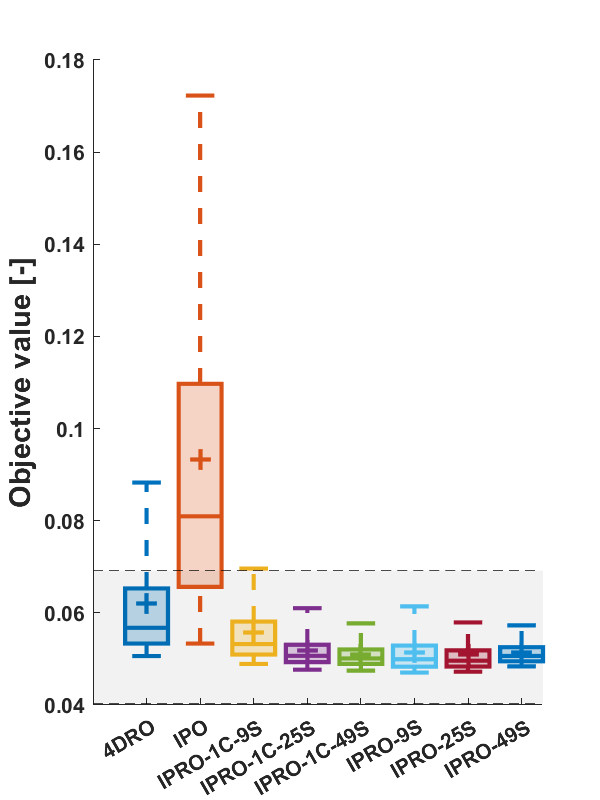}
}}
\makebox[\textwidth][c]{
\subfloat[Case 3a]{
\includegraphics[width=0.4\textwidth]{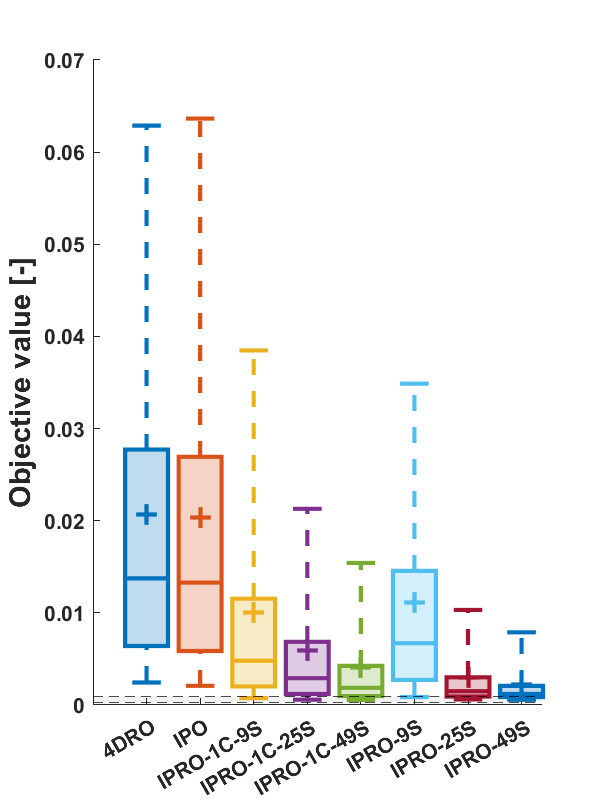}
}
\subfloat[Case 3b]{
\includegraphics[width=0.4\textwidth]{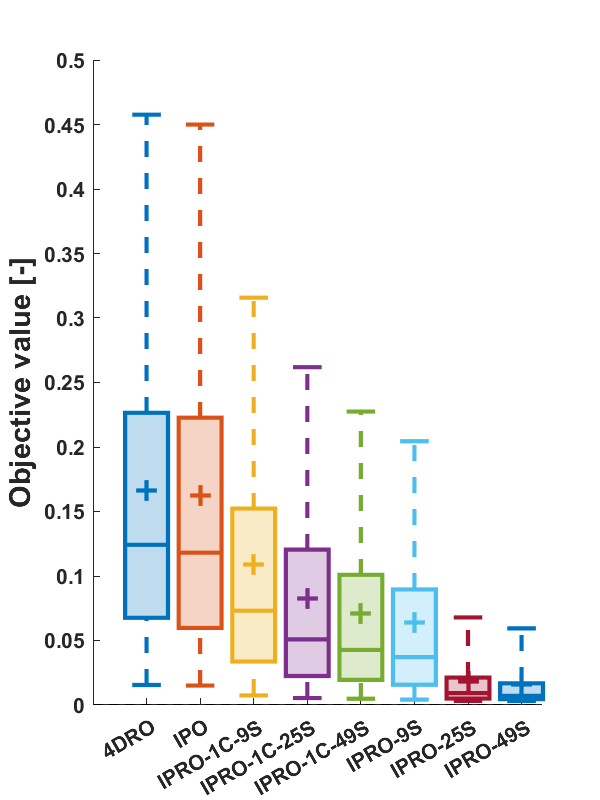}
}}
\caption{The distribution of objective function values over the evaluation motion scenario for each method and case.}
\label{figure_obj_results}
\end{figure}

Figure \ref{figure_obj_results} shows the distribution of objective function values for each method and case.

\printbibliography

\end{document}